\documentclass[final,3p,times,twocolumn]{elsarticle}

\usepackage{amssymb}

\usepackage{url}
\usepackage{hyperref}
\hypersetup{hidelinks}

\journal{Nuclear Inst. and Methods in Physics Research, A}

\begin{document}

    \begin{frontmatter}
    
        
        
        \title{Measurements and TCAD simulations of innovative RSD and DC-RSD LGAD devices for future 4D tracking}
        
        
        \author[a,b]{F.~Moscatelli\corref{cor1}}
            \ead{moscatelli@iom.cnr.it}
            \cortext[cor1]{Corresponding author}
        \author[c,b]{A.~Fondacci}
        \author[d]{R.~Arcidiacono}
        \author[e]{M.~Boscardin}
        \author[f]{N.~Cartiglia}
        \author[e]{M.~Centis Vignali}
        \author[b]{T.~Croci}
        \author[f]{M.~Ferrero}
        \author[e]{O.~Hammad Ali}
        \author[g]{L.~Lanteri}
        \author[g]{A.~Losana}
        \author[g]{L.~Menzio}
        \author[g]{V.~Monaco}
        \author[g]{R.~Mulargia}
        \author[c,b]{D.~Passeri}
        \author[e]{G.~Paternoster}
        \author[g]{F.~Siviero}
        \author[g]{V.~Sola}
        \author[b]{A.~Morozzi}
        
        \affiliation[a]{organization={CNR-IOM of Perugia},
                    addressline={via Pascoli 1}, 
                    city={Perugia},
                    postcode={06123}, 
                    country={Italy}}
        \affiliation[b]{organization={INFN of Perugia},
                    addressline={via Pascoli 1}, 
                    city={Perugia},
                    postcode={06123},
                    country={Italy}}
        \affiliation[c]{organization={Department of Engineering – University of Perugia},
                    addressline={via G. Duranti 93}, 
                    city={Perugia},
                    postcode={06125},
                    country={Italy}}
        \affiliation[d]{organization={Università del Piemonte Orientale},
                    addressline={Largo Donegani 2/3}, 
                    city={Novara},
                    postcode={20100},
                    country={Italy}}
        \affiliation[e]{organization={FBK},
                    addressline={via Sommarive 9}, 
                    city={Trento},
                    postcode={38123},
                    country={Italy}}
        \affiliation[f]{organization={INFN},
                    addressline={via Pietro Giuria}, 
                    city={Torino},
                    postcode={10125},
                    country={Italy}}
        \affiliation[g]{organization={University of Torino},
                    addressline={via Pietro Giuria}, 
                    city={Torino},
                    postcode={10125},
                    country={Italy}}
        
        \begin{abstract}
            This paper summarizes the beam test results obtained with a Resistive Silicon Detector (RSD) (also called AC-Low Gain Avalanche Diode, AC-LGAD) pixel array tested at the DESY beam test facility with a 5 GeV/c electron beam. Furthermore, it describes in detail the simulation results of DC-RSD, an evolution of the RSD design. The simulations campaign described in this paper has been instrumental in the definition of the structures implemented in the Fondazione Bruno Kessler FBK first DC-RSD production. 
                
            The RSD matrix used in this study is part of the second FBK RSD production, RSD2. The best position resolution reached in this test is $\sigma_x = 15$ $\mu m$, about 3.4\% of the pitch. DC-RSD LGAD, are an evolution of the AC-coupled design, eliminating the dielectric and using a DC-coupling to the electronics. The concept of DC-RSD has been finalized using full 3D Technology-CAD simulations of the sensor behavior. TCAD simulations are an excellent tool for designing this innovative class of detectors, enabling the evaluation of different technology options (e.g., the resistivity of the n$^+$ layer, contact materials) and geometrical layouts (shape and distance of the read-out pads). 
        \end{abstract}
        
        
        
    
    \end{frontmatter}
    
    
    \section{Introduction}
    \label{Introduction}
\begin{figure}[htb]
        	\centering
        	\includegraphics[width=0.9\linewidth]{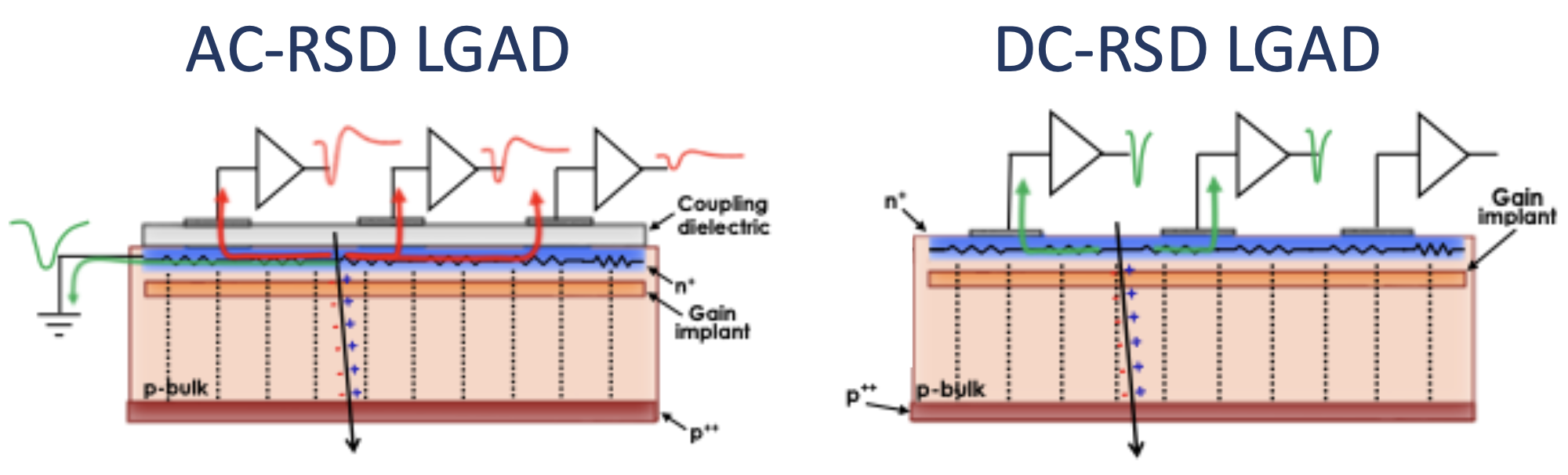}
            \caption{Schematics of a AC-LGAD on the left and of a DC-RSD LGAD on the right.}
            \label{fig:001}
        \end{figure}

        In the past few years, the introduction of two innovations in the design of silicon sensors, (i) Low-Gain Avalanche Diode (LGAD) \cite{1476957,Pellegrini:2014lki} and (ii) Resistive read-out, has significantly improved the performance capabilities of silicon sensors in both spatial and temporal resolutions. Resistive Silicon Detector (RSD) \cite{8846722} is an LGAD (Fig.~\ref{fig:001} left) optimized for both spatial and time measurements (4D-tracking), which make use of an AC-coupled read-out electronic chip and a continuous gain implant without any segmentation. It combines the excellent timing capabilities of LGAD with remarkable spatial resolution, making it an emerging technology for 4D tracking. The major peculiarity of the RSD paradigm is the use of a continuous n$^+$-resistive electrode and a continuous p$^+$-gain layer, allowing a fill factor of 100\%, in contrast to the standard LGAD, which is affected by the lack of gain in the inter-pad region \cite{9060007}. On the other hand, different drawbacks are linked to the nature of RSD paradigm, e.g. the bipolar behavior of the signals due to the use of the AC-coupled read-out, the baseline fluctuation due to the collection of the leakage current at the edge of the sensor, and the position-dependent resolution. The DC-coupled RSD – or DC-RSD (Fig.~\ref{fig:001} right), enables to cope with these issues by eliminating the dielectric and using a DC-coupling to the electronics, while maintaining the advantage of 100\% fill factor.

        Silicon sensors based on resistive read-out \cite{8846722,Tornago:2020otn,Cartiglia:2023izc} combine many of the features needed by future experiments: (i) excellent spatial and temporal resolutions, (ii) low material budget (the active part can be a few tens of $\mu m$ thick), (iii) 100\% fill factor, and (iv) good radiation resistance (presently, up to $1-2\times10^{15}$ $n_{eq}/cm^2$). In addition, given the large pixel size, RSDs are an enabling technology for constructing 4D silicon trackers \cite{Cartiglia:2022ysm} with limited power consumption as they reduce the number of read-out amplifiers by more than an order of magnitude. The benefits of resistive read-out are maximized when the electrode metal area is minimized and shaped to limit the spread of the signal, as reported in a study using a high-precision Transient Current Technique (TCT) set-up \cite{Arcidiacono:2022dsi} to mimic the passage of particles in the sensor. The concept of DC-RSD has been investigated by means of full 3D Technology CAD (TCAD) simulations, evaluating different technology options (e.g., the resistivity of the n$^+$ layer, contact materials) and geometrical layouts (shape and distance of the read-out pads). In particular, a full 3D simulation domain guarantees a very accurate evaluation of the electrical behavior while providing very precise timing information, gaining access to the response of the detector device in terms of conduction and displacement currents. In the second part of this paper, the latest simulation outcome is reported, which has been instrumental for the definition of the design technical implementation for the first DC-RSD production.

    \section{Experimental results}
    \label{ Experimental results}
        \subsection{The beam test experimental setup}
        \label{The Test beam Experimental setup}
\begin{figure}[htb]
            	\centering
            	\includegraphics[width=0.8\linewidth]{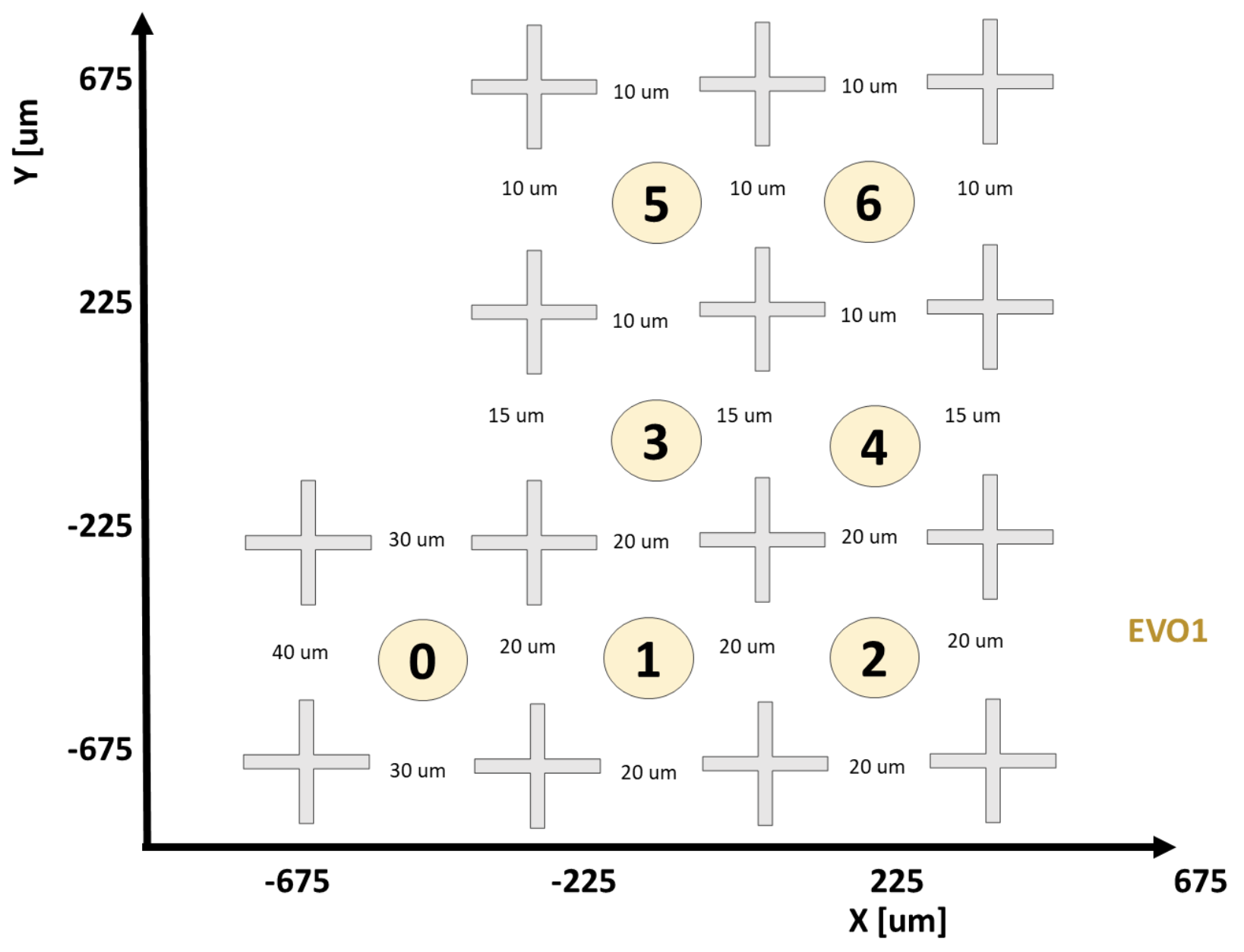}
                \caption{A sketch of the electrodes, the pixels, the gap between the metal arms, and the x-y reference system used in the analysis.}
                \label{fig:002}
            \end{figure}

            The beam test campaign was carried out at the DESY site at Hamburg-Bahrenfeld \cite{Diener:2018qap} with a 5 GeV/c electron beam. The sensor used in the test is a 6$\times$6 matrix of electrodes with a 450 $\mu m$ pitch. The electrodes are cross-shaped, with arms extending in the x and y directions, leaving a small gap between two adjacent arms. The gap length varies from 10 to 40 $\mu m$, while the width of the arm is fixed at 20 $\mu m$. They were wire-bonded to a FAST2 Board \cite{Olave:2021afu,Rojas:2021uex}, a 16-channel amplifier fully custom ASIC developed by INFN Torino using 110 $nm$ CMOS technology. Fig.~\ref{fig:002} shows a sketch of the electrodes, the pixels, the gap between the metal arms, and the x-y reference system used in the analysis. The RSD2 sensor and the FAST2 ASIC were mounted on a custom PCB board. 
            
        \subsection{DESY test beam results}
        \label{DESY test beam results}
            The determination of the hit position in RSD is achieved by first identifying the pixel with the largest signal, and then combining the information from four electrodes placed at the corner of selected pixel. The best hit position reconstruction was achieved using a template of how the signal is shared among the 4 electrodes as a function of the hit position in the pixel. The signal-sharing fraction among the four electrodes as a function of position was measured using data collected at a test beam, creating a so-called ‘‘sharing template’’. More details can be found in \cite{Menzio:2024khz}. For each event, the measured sharing is compared with the sharing templates, and the position in the template that best reproduced the measured sharing is defined as the hit position. 

            \begin{figure}[b]
            	\centering
            	\includegraphics[width=0.8\linewidth]{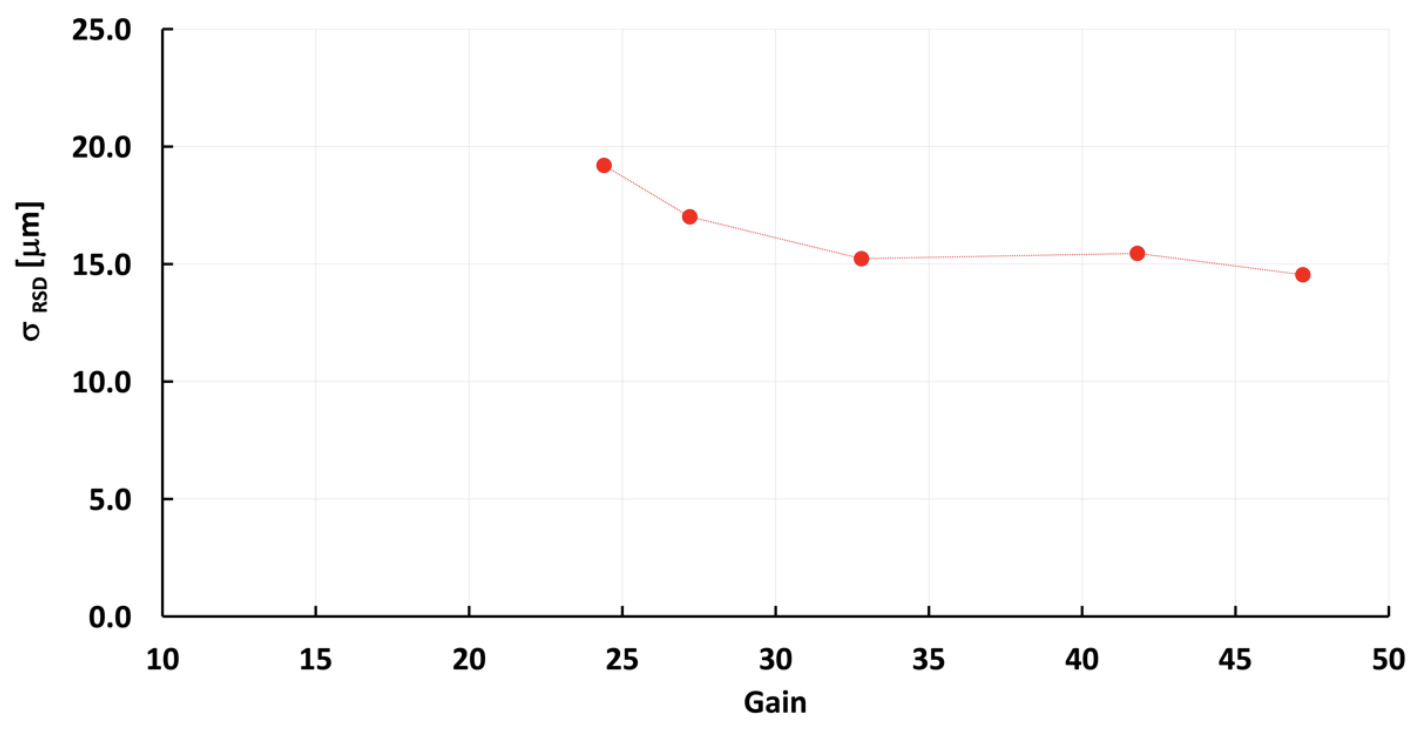}
                \caption{Spatial resolution as a function of the sensor gain as measured at the DESY test beam.}
                \label{fig:003}
            \end{figure}
            
            The RSD spatial resolution is obtained by comparing the RSD hit position with the information provided by the beam line tracking telescope. Figure~\ref{fig:003} reports the results after subtraction in quadrature of $\sigma_{telescope} = 8$~$\mu m$. The results clearly show the excellent resolution that characterizes sensors with resistive readout, better than 4\% of the pitch for 450-micron pixel. The resolution is below 20 $\mu m$ even at the lowest gain.
            
\begin{figure}[htb]
            	\centering
            	\includegraphics[width=0.8\linewidth]{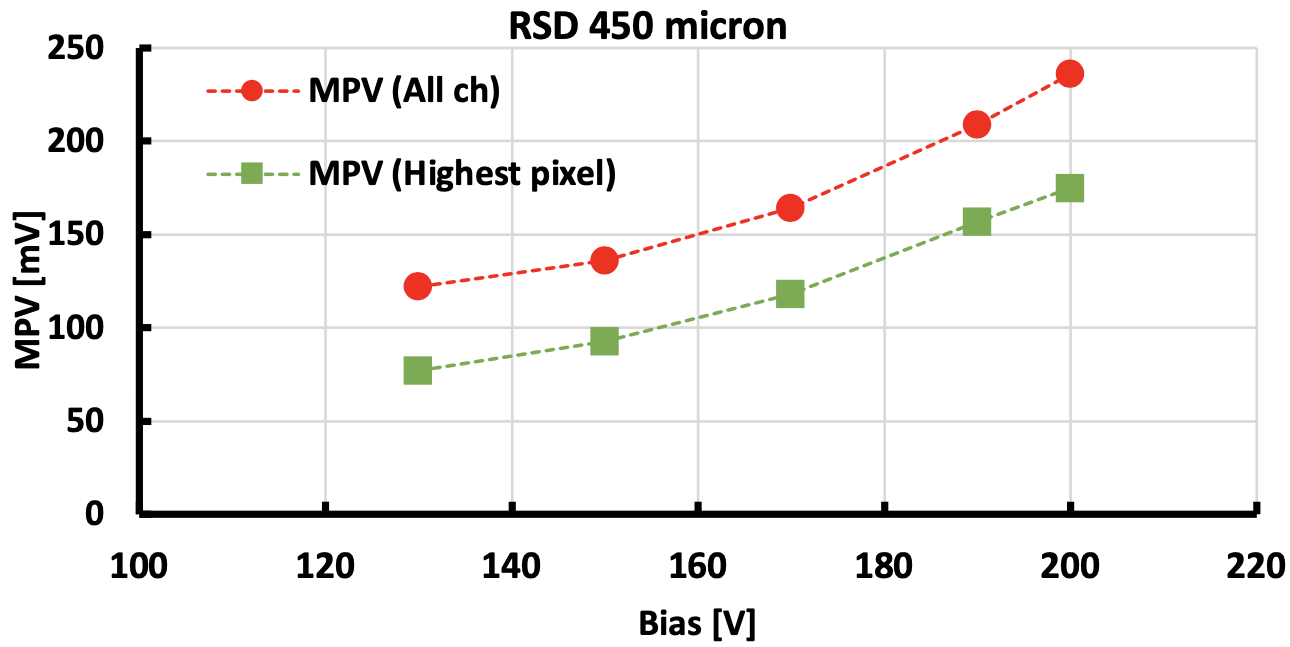}
                \caption{Comparison of Most probable Value (MPV) for RSD 450 micron between the signal collected with all channels and the highest pixel.}
                \label{fig:004}
            \end{figure}

            The results of this beam test show that the existing RSDs have unprecedented performance in terms of spatial resolution. On the other hand, different drawbacks are linked to the nature of the RSD paradigm, which are (i) the bipolar behavior of the generated signals due to the use of the AC-coupled readout; (ii) the baseline fluctuation due to the collection of the leakage current only at the edge of the sensor; (iii) a position dependent resolution; and (iv) the difficulty to scale towards large area sensors. Moreover, signal spreading may involve a larger ($>4$) and variable number of electrodes, leading to a slight deterioration of spatial resolution, which also becomes position-dependent. Performance of these devices with cross shape electrodes is computed using only four electrodes, method which leads to the best results. On average 30\% of the signal leaks outside the area is read by the four electrodes, as shown in Fig.~\ref{fig:004}. The resolution should improve with the full containment of the signal in a predetermined area and the use of DC-RSD device can improve the performances and scalability to large devices. Moreover, as said in the introduction, by eliminating the dielectric, and hence by performing a DC-coupled read-out, it is possible to have unipolar signals, absence of baseline fluctuation, a well-controlled charge sharing, and large sensitive areas (in the order of centimeters), while maintaining the advantages of signal spreading and 100\% fill factor.
            
    \section{TCAD optimization of DC-RSD}
    \label{TCAD optimization of DC-RSD}
\begin{figure}[hb]
        	\centering
        	\includegraphics[width=0.9\linewidth]{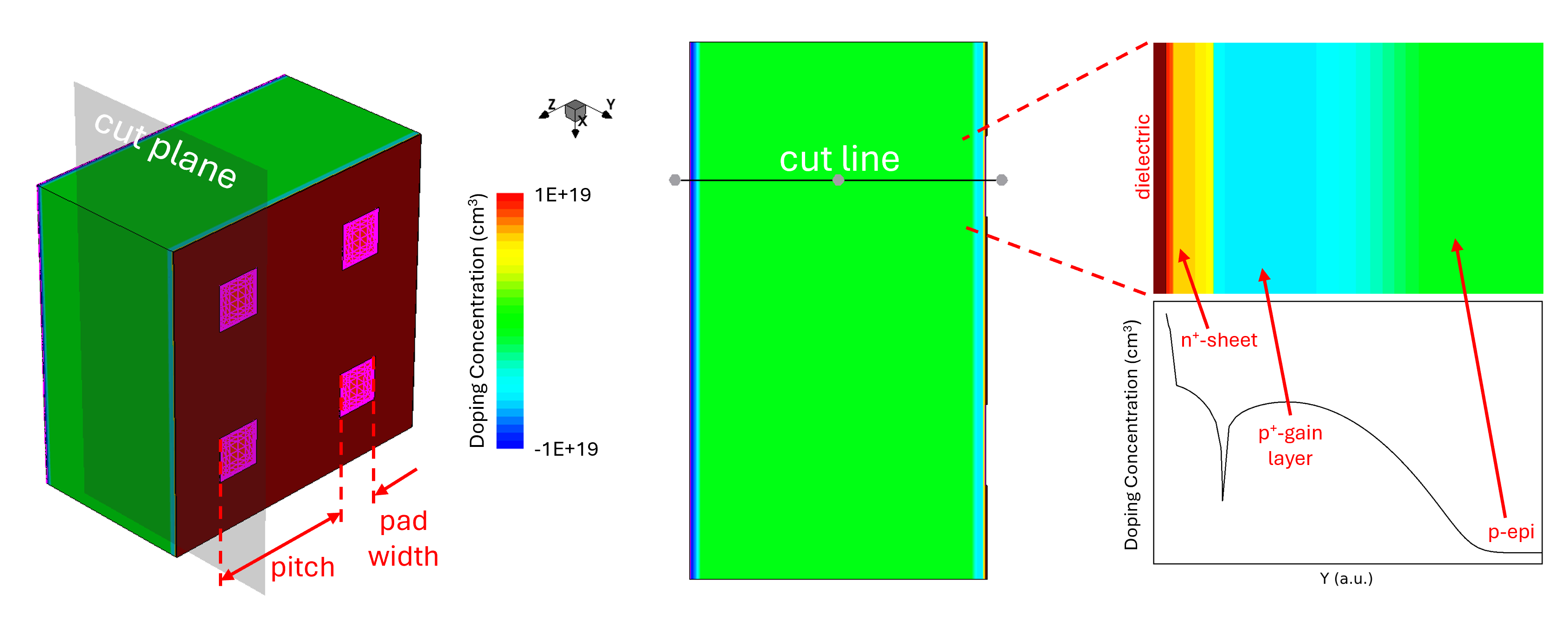}
            \caption{On the left, TCAD model of a generic four-pad DC-RSD device in the 3-D domain, used for the first analysis \cite{10411098}. In the center, 2D cross section of the simulated domain is shown. On the right, X-Y cut of the doping profile is reported, where the n$^+$-resistive sheet, the p$^+$-gain layer implants and the p-epi are visible.}
            \label{fig:005}
        \end{figure}

        State-of-the-art TCAD simulation software represents a sophisticated tool for the design, and analysis of semiconductor detectors aiming at their performance optimization. The concept of DC-RSD has been investigated within the Synopsys Sentaurus TCAD environment \cite{tcad} using full 3D TCAD simulations, enabling the evaluation of different technology options (e.g., the resistivity of the n$^+$ layer, contact materials) and geometrical layouts (i.e. shape and distance of the read-out pads). A full 3D simulation domain guarantees a very accurate assessment of the electrical behavior while providing very precise timing information, gaining access to the response of the detector device in terms of currents. Several challenges have to be addressed for the simulation of DC-RSD, due to the nature of its design and the large pixel size of the order of a few millimeters. To optimize the simulation domain without losing accuracy, the simulated layout comprehends 2$\times$2 electrodes (Fig.~\ref{fig:005}), and a border has been considered thus avoiding eventual boundary effects. The mesh size is in the order of  290 k-points and the simulation time for a Minimum Ionizing Particle (MIP) impinging the structure is in the order of 20 hours (using a workstation with 16 CPUs). Full-3D TCAD simulations have been carried out in order to analyze the device behavior in terms of electrical properties and response after the passage of a charged particle. A p$^+$-gain layer has been realized in the first few microns from the surface. By varying the shape of the n$^+$-resistive sheet implant mainly in terms of thickness and doping concentration, we have explored the impact of different values of the sheet resistance ($R_{s,n^+}$) on the steady state (DC) and transient behavior of the sensor. The results of this first analysis have been published in \cite{10411098}. These first detailed 3D TCAD simulations allowed for the identification of a set of sensor design parameters for the first production, thus optimizing the sensor performances and its layout, avoiding devices that would not perform well. We explored an extensive range of values for the sheet resistance, aiming at optimizing the reconstruction of the particle impact point coordinates, the isolation between pads, and the timing capabilities. The best results have been obtained with a sheet resistance $R_{sheet}$ in the range between 1 and 2 k$\Omega$/sq. 
        
        Another parameter analyzed with this four-pad DC-RSD structure, is the pitch size. The simulation results show that with a constant pad width, an increase in pitch size corresponds to a heightened degree of distortion in reconstructing impact points using the Center-of-Gravity method \cite{10411098}. A reasonable lower limit for the pitch size to have a good compromise between the reconstruction of the particle impact positions and the number of readout channels is 100 $\mu m$.
        
        \begin{figure}[ht]
        	\centering
        	\includegraphics[width=0.8\linewidth]{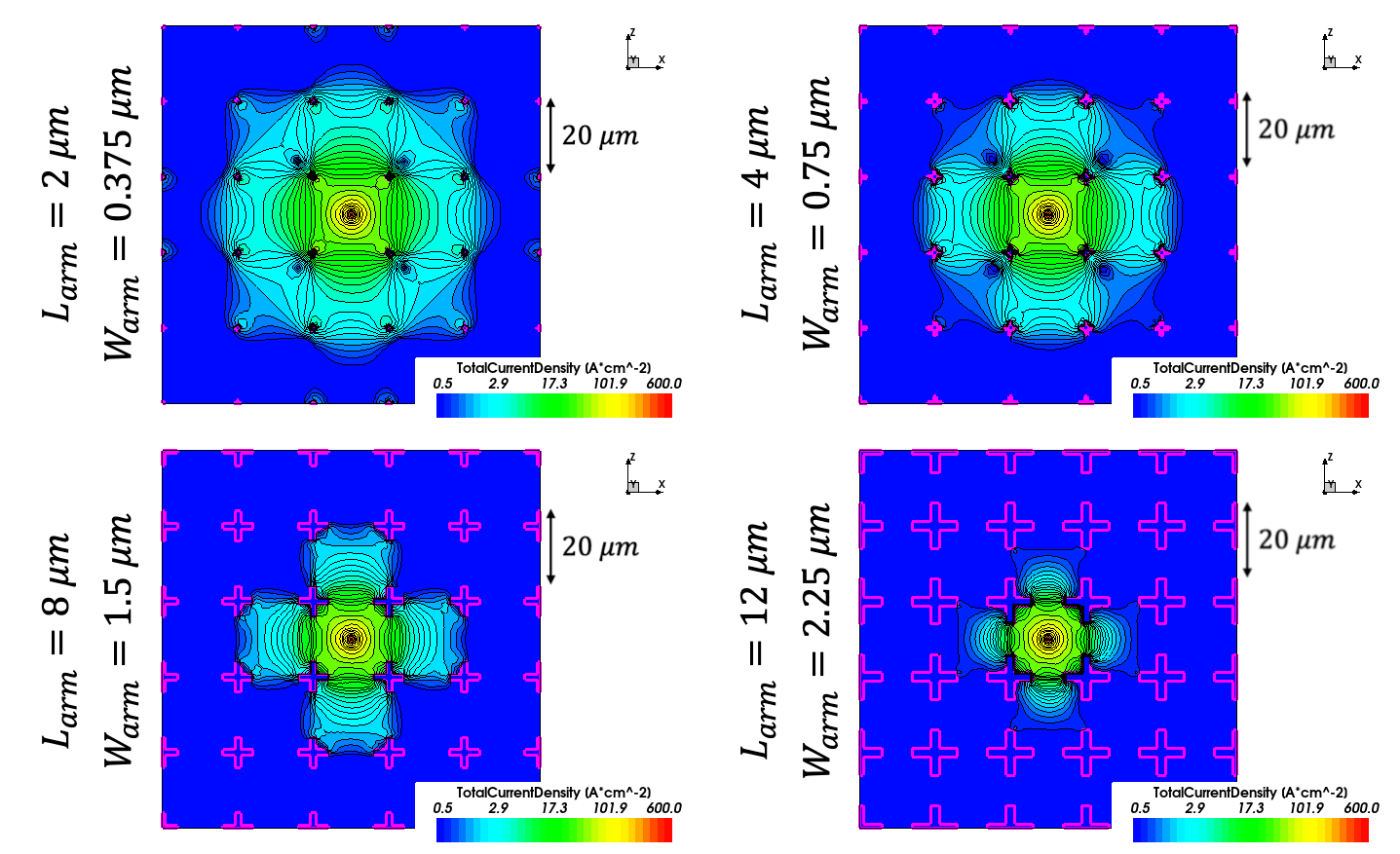}
            \caption{Spatial distribution of electron current over the detector surface of four 20 $\mu m$-pitch 5$\times$5 matrix structure DC-RSD detectors with different width and length of the cross pads.}
            \label{fig:006}
        \end{figure}
        
        A second DC-RSD structure in the 3D domain has been simulated considering a 5$\times$5 pixel matrix structure with fixed pitch at 20 $\mu m$, aiming at analyzing the signal confinement. The sheet resistance of the n$^+$ implanted layer has been considered of $\cong\!2$ k$\Omega$/sq, the contact resistance between the metal and the n$^+$ layer has been set to 10 $\Omega$. Simulations have been carried out at 300~K, considering Massey avalanche model and applying a substrate voltage of -200 V. Different size for the cross- shaped pads were considered. Fig.~\ref{fig:006} shows the current density maps after the crossing of a MIP in the center of the structure for four different length and width of the cross pads. The signal is well-confined inside the cell, when the electrodes are a significant fraction of the pitch.  
        \begin{figure}[hb]
        	\centering
        	\includegraphics[width=0.8\linewidth]{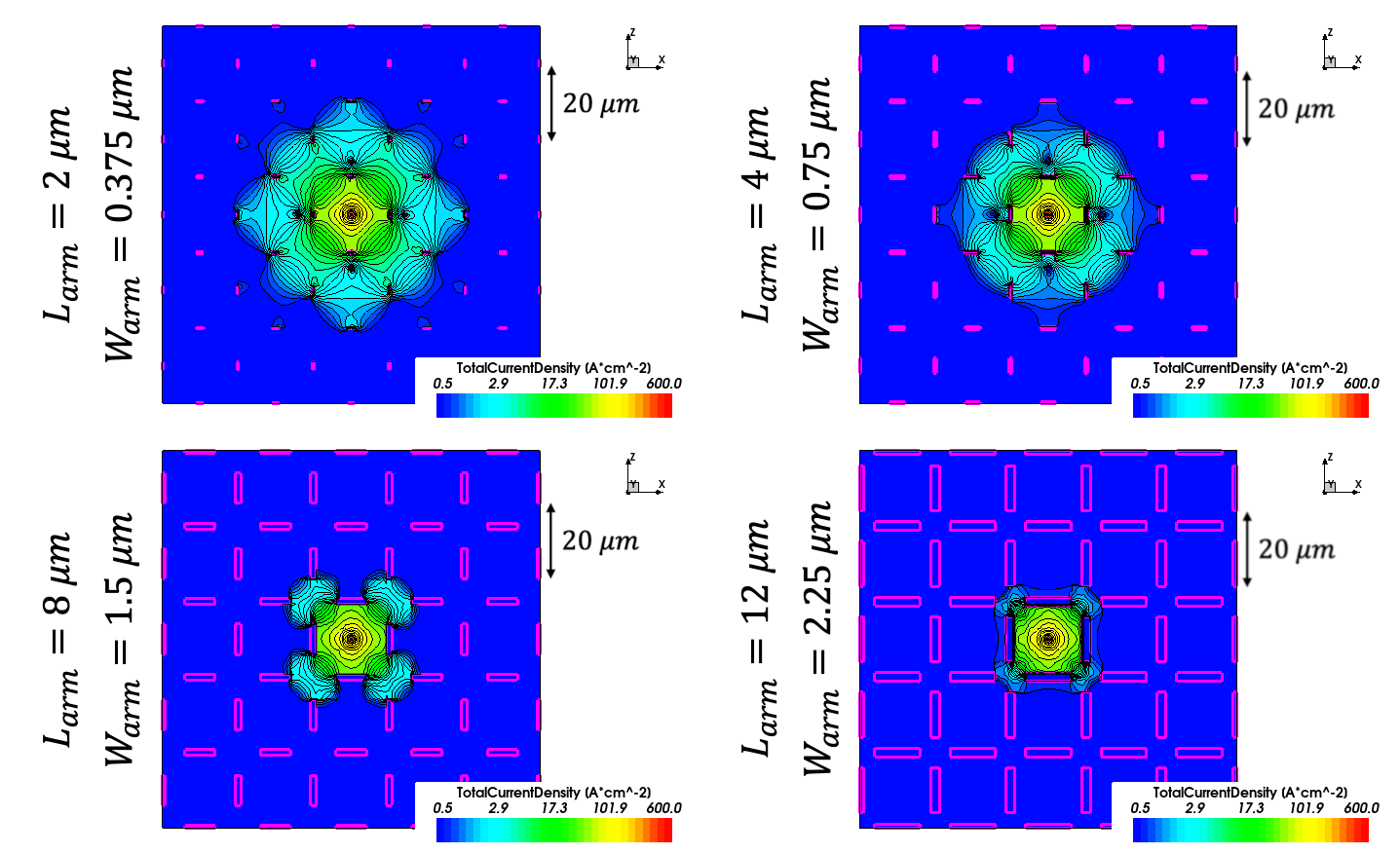}
            \caption{Spatial distribution of electron current over the detector surface of four 20 $\mu m$-pitch 5$\times$5 matrix structure DC-RSD detectors with different width and length of the bar-shaped pads.}
            \label{fig:007}
        \end{figure}
        
        A similar conclusion can be extended to bar-shaped pads as shown in Fig.~\ref{fig:007}, where the investigation of the signal confinement as a function of the pad size is shown. Both for cross and bar pads, using the longer electrodes, it is possible to collect 96-97\% of the injected charge in the central cell. Cross-shaped pads configuration allow to obtain this result with fewer electrodes (36 vs 60 in this 5$\times$5 matrix structure). 
        
\begin{figure}[ht]
        	\centering
        	\includegraphics[width=0.65\linewidth]{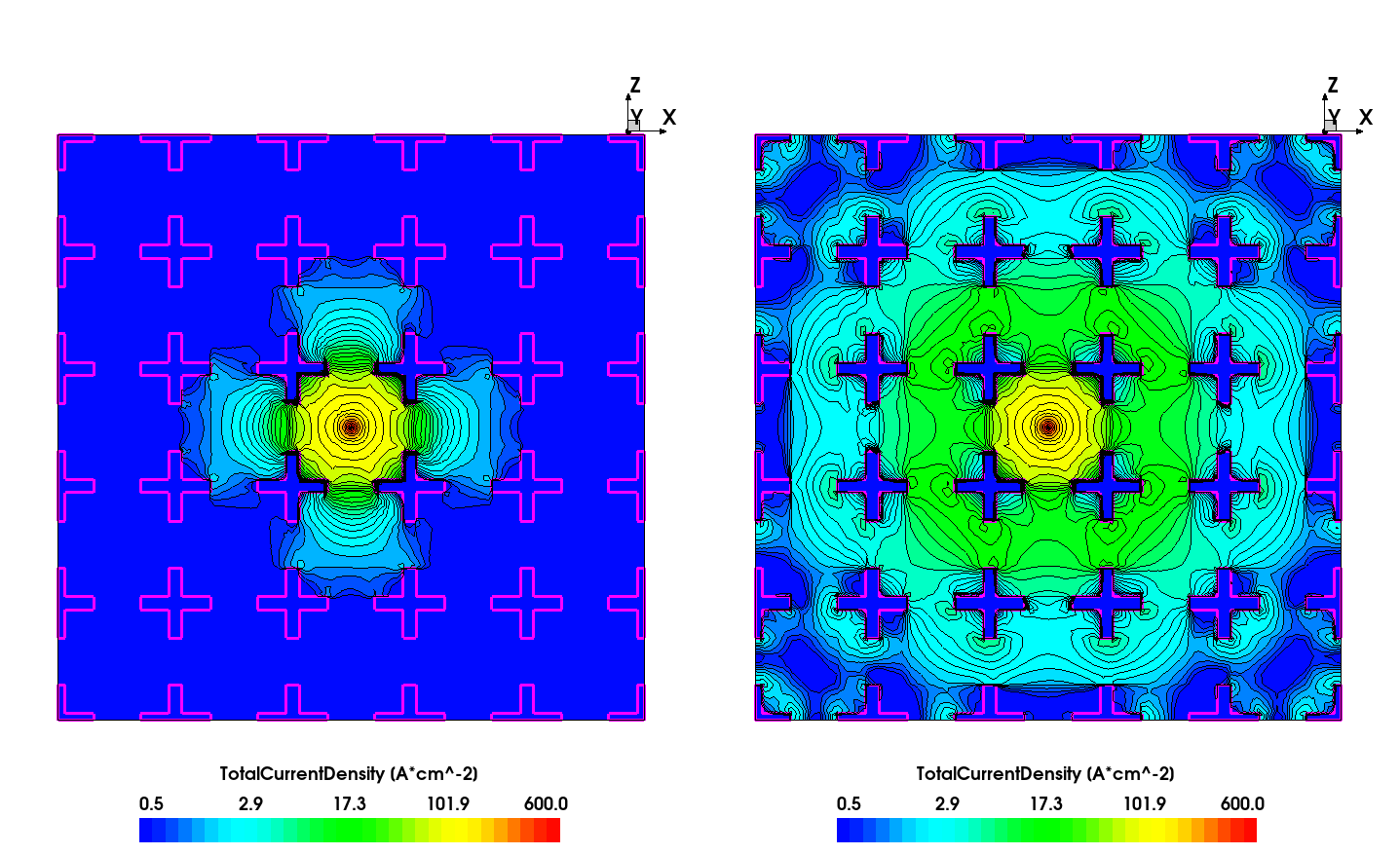}
            \caption{Spatial distribution of electron current over the detector surface of a 20 $\mu m$-pitch 5$\times$5 matrix structure DC-RSD detector with contact resistance 10 $\Omega$ (left) and 1 k$\Omega$ (right), a few hundred picoseconds after the passage of the MIP.}
            \label{fig:008}
        \end{figure}

        Another crucial aspect is the contact resistance between the metal pad and the n$^+$ layer. As shown in Fig.~\ref{fig:008} if the contact resistance is very low (i.e., 10 $\Omega$) the signal is well confined in the central cell. In contrast, if the contact resistance is high (i.e., 1 k$\Omega$) the charges find a lower resistance path in the n$^+$ layer, making it impossible to confine the signal.
        
\begin{figure}[hb]
        	\centering
        	\includegraphics[width=0.65\linewidth]{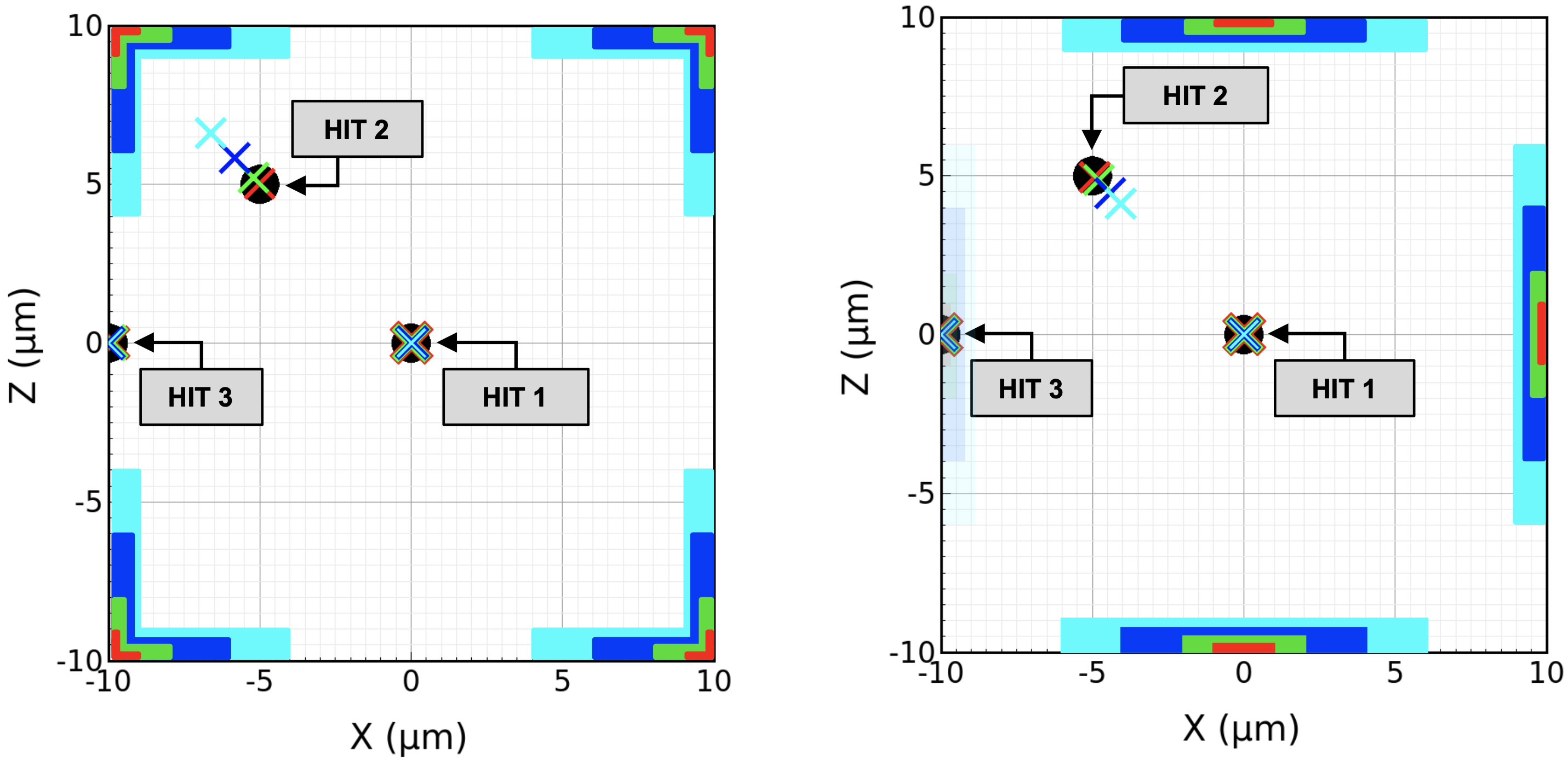}
            \caption{Map of the injected (black dot) and reconstructed (x symbols) impact positions, which has been obtained by TCAD simulation of a 20 $\mu m$-pitch four-pad DC-RSD detector with cross pads (left) and bar-shaped pads (right). Different colors are related to different dimensions of the pads.}
            \label{fig:009}
        \end{figure}

        Fig.~\ref{fig:009} shows the map of the injected (black dot) and reconstructed (x symbols) impact positions, which has been obtained by TCAD simulation of a 20 $\mu m$-pitch four-pad DC-RSD detector using cross pads (left) or bar-shaped pads (right). As shown in Fig.~\ref{fig:009}, the larger the pad dimensions the greater the level of distortion in reconstructing the impact position coordinates. Moreover, when a particle strikes an electrode away from its center, the information about the impact position is altered (located in the center). Therefore, the implementation of electrodes with long arms is not suitable.
        
\begin{figure}[htb]
        	\centering
        	\includegraphics[width=0.9\linewidth]{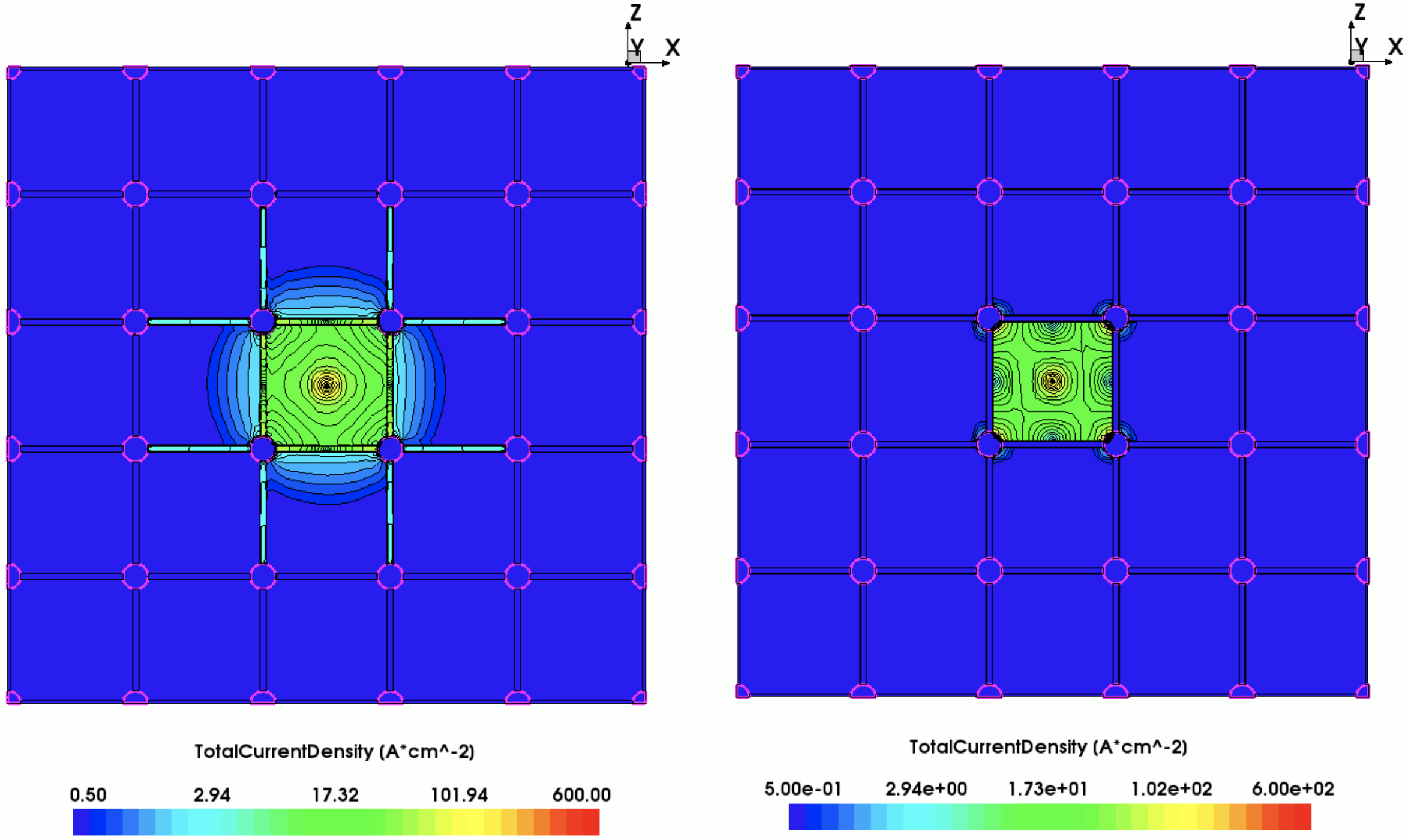}
            \caption{Spatial distribution of electron current over the detector surface of a 20 $\mu m$-pitch 5$\times$5 matrix structure DC-RSD detector with strip resistors (a) and with trenches (b).}
            \label{fig:010}
        \end{figure}

        To obtain a very good confinement of the signal, avoiding the distortion of the hit position reconstruction, two different solutions have been analyzed. The first one (Fig.~\ref{fig:010}a) is the use of inter-pad resistors. We investigated different values of the strip resistance by varying both the geometry and the material. Considering that this resistance must be sufficiently high to prevent to short-circuiting the front-end electronics, we obtained a very good signal confinement across a broad range of values (2-40\% of the n$^+$ sheet resistance).
        A second strategy is the use of pixels with isolating trenches. Trenches have been well manufactured and characterized for the production of Silicon Photomultiplier (SiPM) or Trench-isolated LGAD \cite{Paternoster:2020qkd}. 
        As shown in Fig.~\ref{fig:010}a and~\ref{fig:010}b in both cases the charge is, in turn, collected almost entirely by the four pads of the affected pixel.
        
    \section{Conclusion}
    \label{Conclusion}
        This paper reports on the spatial resolution of an RSD 450 $\mu m$ pitch pixel array, demonstrating the effectiveness of RSD sensors in achieving concurrent excellent position resolution. The sensor matrix used in this study is part of the second Fondazione Bruno Kessler (FBK) RSD production (RSD2), consisting of seven 450 $\mu m$ pitch pixels with cross-shaped electrodes, covering an area of about 1.5 mm$^2$. The electrodes were read out by the FAST2 ASIC, a 16-channel amplifier fully custom ASIC developed by INFN Torino using 110 $nm$ CMOS technology. The study was performed at the DESY beam test facility with a 5 GeV/c electron beam. Key findings include achieving a position resolution of $\sigma_x = 15$ $\mu m$ , approximately 3.5\% of the pitch. The existing resistive read-out LGAD sensors (AC-coupled RSD) demonstrate unprecedented performance in terms of spatial resolution. This innovative sensor concept looks very promising for the future 4D-tracking detectors. The DC-coupled version of the RSD should provide improved performance and scalability to larger devices. Detailed 3D TCAD simulations allowed for the identification of a set of sensor design parameters for the first production, optimizing the layout, avoiding devices that would not perform well. To avoid introducing distortions in the reconstruction of the impact position, the use of small electrodes is suitable. Isolating trenches, interrupting the resistive n$^+$ layer, excellently confines the signal. Resistive strips are also good at confining the signal and there is a high tolerance for the manufacturing process. Considering the simulation results, devices with squared matrices of electrodes (dot-like), without and with isolating trenches, will be implemented in the first batch of DC-RSD devices.
    
    \section{Acknowledgments}
    \label{Acknowledgments}
        The measurements leading to these results have been performed at the test beam facility at DESY Hamburg (Germany), a member of the Helmholtz Association (HGF). This project has received funding from the European Union Horizon Europe research and innovation programme under grant agreement No 101057511. This project has received funding from the European Union’s Horizon 2020 research and innovation programme under GA No 101004761, from the Italian MIUR PRIN under GA No 2017L2XKTJ and GA No 2022KLK4LB and from INFN - CSN V - 4DSHARE project.

    
    
     \bibliographystyle{elsarticle-num} 
     \bibliography{cas-refs}

\begin{thebibliography}{10}
\expandafter\ifx\csname url\endcsname\relax
  \def\url#1{\texttt{#1}}\fi
\expandafter\ifx\csname urlprefix\endcsname\relax\def\urlprefix{URL }\fi
\expandafter\ifx\csname href\endcsname\relax
  \def\href#1#2{#2} \def\path#1{#1}\fi

\bibitem{1476957}
J.~Conradi, The distribution of gains in uniformly multiplying avalanche photodiodes: Experimental, IEEE Transactions on Electron Devices 19~(6) (1972) 713--718.
\newblock \href {https://doi.org/10.1109/T-ED.1972.17486} {\path{doi:10.1109/T-ED.1972.17486}}.

\bibitem{Pellegrini:2014lki}
G.~Pellegrini, et~al., {Technology developments and first measurements of Low Gain Avalanche Detectors (LGAD) for high energy physics applications}, Nucl. Instrum. Meth. A 765 (2014) 12--16.
\newblock \href {https://doi.org/10.1016/j.nima.2014.06.008} {\path{doi:10.1016/j.nima.2014.06.008}}.

\bibitem{8846722}
M.~Mandurrino, R.~Arcidiacono, M.~Boscardin, N.~Cartiglia, G.~F. Dalla~Betta, M.~Ferrero, F.~Ficorella, L.~Pancheri, G.~Paternoster, F.~Siviero, M.~Tornago, Demonstration of 200-, 100-, and 50- $\mu m$ pitch resistive ac-coupled silicon detectors (rsd) with 100
\newblock \href {https://doi.org/10.1109/LED.2019.2943242} {\path{doi:10.1109/LED.2019.2943242}}.

\bibitem{9060007}
M.~Mandurrino, R.~Arcidiacono, M.~Boscardin, N.~Cartiglia, G.-F.~D. Betta, M.~Ferrero, F.~Ficorella, L.~Pancheri, G.~Paternoster, F.~Siviero, M.~Tornago, First production of 50-$\mu m$-thick resistive ac-coupled silicon detectors (rsd) at fbk, in: 2019 IEEE Nuclear Science Symposium and Medical Imaging Conference (NSS/MIC), 2019, pp. 1--3.
\newblock \href {https://doi.org/10.1109/NSS/MIC42101.2019.9060007} {\path{doi:10.1109/NSS/MIC42101.2019.9060007}}.

\bibitem{Tornago:2020otn}
M.~Tornago, et~al., {Resistive AC-Coupled Silicon Detectors: principles of operation and first results from a combined analysis of beam test and laser data}, Nucl. Instrum. Meth. A 1003 (2021) 165319.
\newblock \href {http://arxiv.org/abs/2007.09528} {\path{arXiv:2007.09528}}, \href {https://doi.org/10.1016/j.nima.2021.165319} {\path{doi:10.1016/j.nima.2021.165319}}.

\bibitem{Cartiglia:2023izc}
N.~Cartiglia, et~al., {Resistive Read-out in Thin Silicon Sensors with Internal Gain} (1 2023).
\newblock \href {http://arxiv.org/abs/2301.02968} {\path{arXiv:2301.02968}}.

\bibitem{Cartiglia:2022ysm}
N.~Cartiglia, R.~Arcidiacono, M.~Costa, M.~Ferrero, G.~Gioachin, M.~Mandurrino, L.~Menzio, F.~Siviero, V.~Sola, M.~Tornago, {4D tracking: present status and perspectives}, Nucl. Instrum. Meth. A 1040 (2022) 167228.
\newblock \href {http://arxiv.org/abs/2204.06536} {\path{arXiv:2204.06536}}, \href {https://doi.org/10.1016/j.nima.2022.167228} {\path{doi:10.1016/j.nima.2022.167228}}.

\bibitem{Arcidiacono:2022dsi}
R.~Arcidiacono, et~al., {High-precision 4D tracking with large pixels using thin resistive silicon detectors}, Nucl. Instrum. Meth. A 1057 (2023) 168671.
\newblock \href {http://arxiv.org/abs/2211.13809} {\path{arXiv:2211.13809}}, \href {https://doi.org/10.1016/j.nima.2023.168671} {\path{doi:10.1016/j.nima.2023.168671}}.

\bibitem{Diener:2018qap}
R.~Diener, et~al., {The DESY II Test Beam Facility}, Nucl. Instrum. Meth. A 922 (2019) 265--286.
\newblock \href {http://arxiv.org/abs/1807.09328} {\path{arXiv:1807.09328}}, \href {https://doi.org/10.1016/j.nima.2018.11.133} {\path{doi:10.1016/j.nima.2018.11.133}}.

\bibitem{Olave:2021afu}
E.~J. Olave, F.~Fausti, N.~Cartiglia, R.~Arcidiacono, H.~F.~W. Sadrozinski, A.~Seiden, {Design and characterization of the FAST chip: a front-end for 4D tracking systems based on Ultra-Fast Silicon Detectors aiming at 30 ps time resolution}, Nucl. Instrum. Meth. A 985 (2021) 164615.
\newblock \href {https://doi.org/10.1016/j.nima.2020.164615} {\path{doi:10.1016/j.nima.2020.164615}}.

\bibitem{Rojas:2021uex}
A.~M. Rojas, et~al., {Amplifier-Discriminator ASICs to Read Out Thin Ultra-Fast Silicon Detectors for ps Resolution}, in: {2021 IEEE Nuclear Science Symposium (NSS) and Medical Imaging Conference (MIC) and 28th International Symposium on Room-Temperature Semiconductor Detectors}, 2021.
\newblock \href {https://doi.org/10.1109/NSS/MIC44867.2021.9875441} {\path{doi:10.1109/NSS/MIC44867.2021.9875441}}.

\bibitem{Menzio:2024khz}
L.~Menzio, et~al., {First test beam measurement of the 4D resolution of an RSD 450 microns pitch pixel matrix connected to a FAST2 ASIC} (2 2024).
\newblock \href {http://arxiv.org/abs/2402.01517} {\path{arXiv:2402.01517}}.

\bibitem{tcad}
Synopsys sentaurus tcad (online), \url{https://synopsys.com}.

\bibitem{10411098}
T.~Croci, L.~Menzio, R.~Arcidiacono, M.~Arneodo, P.~Asenov, N.~Cartiglia, M.~Ferrero, A.~Fondacci, V.~Monaco, A.~Morozzi, F.~Moscatelli, R.~Mulargia, E.~Robutti, V.~Sola, D.~Passeri, A two-prong approach to the simulation of dc-rsd: Tcad and spice, IEEE Transactions on Nuclear Science 71~(2) (2024) 127--134.
\newblock \href {https://doi.org/10.1109/TNS.2024.3356826} {\path{doi:10.1109/TNS.2024.3356826}}.

\bibitem{Paternoster:2020qkd}
G.~Paternoster, G.~Borghi, M.~Boscardin, N.~Cartiglia, M.~Ferrero, F.~Ficorella, F.~Siviero, A.~Gola, P.~Bellutti, {Trench-Isolated Low Gain Avalanche Diodes (TI-LGADs)}, IEEE Electron. Dev. Lett. 41~(6) (2020) 884--887.
\newblock \href {https://doi.org/10.1109/LED.2020.2991351} {\path{doi:10.1109/LED.2020.2991351}}.

\end{thebibliography}
    
    
    
    
    
\end{document}